\documentstyle[12pt]{article}
\begin{document}
\begin{center}{Physical Review D 60, 117503 (1999)}\\[1cm]\end{center} 
\begin{center}{\large Are the reactions $\gamma\gamma\to VV'$ a challenge for
the factorized Pomeron\\[0.2cm] at high energies\,?}\\[2cm]
{\large N.N. Achasov and G.N. Shestakov}\\[0.5cm] {\it Laboratory of 
Theoretical Physics,\\ S.L. Sobolev Institute for Mathematics,\\
630090, Novosibirsk 90, Russia}\\[2.5cm] Abstract\\[1cm] \end{center}
We would like to point to the strong violation of the putative factorized 
Pomeron exchange model in the reactions $\gamma\gamma\to VV'$ in the 
high-energy region where this model works fairly well in all other 
cases.\\[0.5cm]
PACS number(s): 12.40.Nn, 13.60.Le, 13.65.+i
\newpage
The factorized Pomeron exchange model is one of the most well-grounded and good
working phenomenological models in high energy physics. Currently this model is
particularly used in analyses of the DESY $ep$ collider HERA and CERN $e^+e^-$
collider LEP2 data on $\gamma p$ and $\gamma\gamma$ interactions (see, for 
example, Refs. [1-5]).

About five years ago, immediately after the ARGUS observation of $\gamma\gamma
\to\rho^0\phi$ [6], we intended to publish a work entitled ``Is the reaction
$\gamma\gamma\to\rho^0\phi$ a challenge for the factorization model at high
energies\,?" As a result, there appeared the paper: ``Estimate of $\sigma(
\gamma\gamma\to VV')$ at high energies" [7], in which, on the basis of the
factorization model, the cross section for the reaction $\gamma\gamma\to\rho^0
\phi$ was estimated in the range $11.5\leq W_{\gamma\gamma}\leq18.4$ GeV (where
$W_{\gamma\gamma}$ is the $\gamma\gamma$ center-of-mass energy): $\sigma(\gamma
\gamma\to\rho^0\phi)=(1.2-2.4)$ nb. We obtained the estimate taking
into account all possible combinations of the existing sets of the data on the
reactions $\gamma p\to\rho^0p$ and $\gamma p\to\phi p$, in the incident photon
laboratory energy range from 70 to 180 GeV, in the factorization relation for
the $\gamma\gamma\to\rho^0\phi$,\, $\gamma p\to\rho^0p$,\, $\gamma p\to\phi p$,
\,and \,$pp\to pp$ cross sections [7].
A comparison of this estimate with the ARGUS data, $\sigma(\gamma\gamma\to\rho
^0\phi)=(0.16\pm0.16)$ nb for $3.25\leq W_{\gamma\gamma}\leq3.5$ GeV, has shown
that between 3.5 and 11.5 GeV the $\gamma\gamma\to\rho^0\phi$ reaction cross
section can increase by an order of magnitude. Nothing of the kind
has yet occurred in elastic and quasielastic reactions with
the Pomeron exchange and with particles involving light quarks. Therefore,
such an unusually strong rise of $\sigma(\gamma\gamma\to\rho^0\phi)$
expected from the factorization model and from the ARGUS data would be
essentially a real challenge for our current ideas about the dynamics of
quasi-two-body reactions. Why is the $\gamma\gamma\to\rho^0\phi$ cross section
so small near 3.5 GeV\,? In Ref. [7] we concluded that either we faced a new
physical phenomenon in the reaction $\gamma\gamma\to\rho^0\phi$
or the ARGUS data [6] were underestimated for some reason.

In Ref. [7] we also applied the factorization model to other reactions $\gamma
\gamma\to VV'$ ($V(V')=\rho^0,\,\omega,\,\phi$). In particular, for the $\rho^0
\rho^0$ and $\rho^0\omega$ channels in the range $11.5\leq W_{\gamma\gamma}\leq
18.4$ GeV, we obtained the following estimates: $\sigma(\gamma\gamma\to\rho^0
\rho^0)=(9.9-21)$ nb and $\sigma(\gamma\gamma\to\rho^0\omega)=(1.9-3.8)$ nb. 
Note that the central values of our estimates for $\sigma(\gamma\gamma\to\rho^0
\rho^0)$,\, $\sigma(\gamma\gamma\to\rho^0\omega)$,\, and \,$\sigma(\gamma\gamma
\to\rho^0\phi)$ are in excellent agreement with the similar ones obtained in 
Ref. [3] for the other purpose.

Here we want once again to question the factorization model for the reactions
$\gamma\gamma\to VV'$ in connection with the imposing data obtained by the
L3 Collaboration on the reaction $\gamma\gamma\to\rho^0\rho^0$, which has been
reported at the International Workshop on $e^+e^-$ Collisions from $\phi$ to
$J/\psi$ in Novosibirsk [8].

Figure 1 shows the cross section for the process $\gamma\gamma\to\pi^+\pi^-\pi
^+\pi^-$ measured by the L3 Collaboration in the energy range from 0.75 to 4.9
GeV [8]. For $W_{\gamma\gamma}<2$ GeV, $\sigma(\gamma\gamma\to\pi^+\pi^-\pi^+
\pi^-)$ is rather large and is strongly dominated by $\rho^0\rho^0$ production
[8,9]. Let us now look at the high $W_{\gamma\gamma}$ 
region. For $4.5\leq W_{\gamma\gamma}\leq4.9$ GeV, as is clear from the L3 data
shown in Fig. 1, $\sigma(\gamma\gamma\to\rho^0\rho^0)$ is certainly less than 
1.5 nb. Thus, for the reaction $\gamma\gamma\to\rho^0\rho^0$ one can repeat
exactly the same statements which have been done in Ref. [7] and mentioned
above in connection with the data on $\rho^0\phi$ production and the
factorization model prediction.

However, we now assess the situation of the factorization model as more
critical. The fact is that the L3 Collaboration has already measured the rate
of $\gamma\gamma\to\rho^0\rho^0$ events up to $W_{\gamma\gamma}=10$ GeV [5].
If the $\gamma\gamma\to\rho^0\rho^0$ cross section does not increase
approximately by an order of magnitude with increasing $W_{\gamma\gamma}$ from
5 to 10 GeV, then it will signify that the factorization model for the reaction
$\gamma\gamma\to\rho^0\rho^0$ is a failure in the energy region where this
works fairly well in other cases.

A failure of the factorization should be expected not only in the $\rho^0\rho^0
$ and $\rho^0\phi$ channels but in the $\rho^0\omega$, \,$\omega\omega$, \,$
\omega\phi$, \,and\, $\phi\phi$ ones, too, because, at high energies, the
reactions $\gamma\gamma\to\rho^0\rho^0$, \,$\gamma\gamma\to\rho^0\omega$, \,$
\gamma\gamma\to\rho^0\phi$, \,$\gamma\gamma\to\omega\omega$, \,$\gamma\gamma\to
\omega\phi$, \,and \,$\gamma\gamma\to\phi\phi$ are due to have similar
mechanisms.

Thus, it may happen
that either the $\gamma\gamma\to\rho^0\rho^0$ reaction cross section reaches
the magnitude expected on the basis of the factorization model only at still
higher energies, and there is a need to look for a specific dynamical reason 
for so defiant a phenomena in the formation mechanism of the Pomeron exchange
for quasi-two-body reactions, or the L3 detection efficiency for the process
$\gamma\gamma\to\rho^0\rho^0$, which is small at high $W_{\gamma\gamma}$ [5],
has been, however, overestimated by an order of magnitude. Both of these 
possibilities are thus extremely important and require an immediate
elucidation. However, it seems almost improbable that the same accident has 
occurred in measuring the two different reactions $\gamma\gamma\to\rho^0\phi$
and $\gamma\gamma\to\rho^0\rho^0$ with the two different detectors ARGUS and 
L3, respectively.

We would like to thank V. Schegelsky for the discussion of the L3 data.
 
\begin{center} \vspace*{1cm} {\bf FIGURE CAPTION}
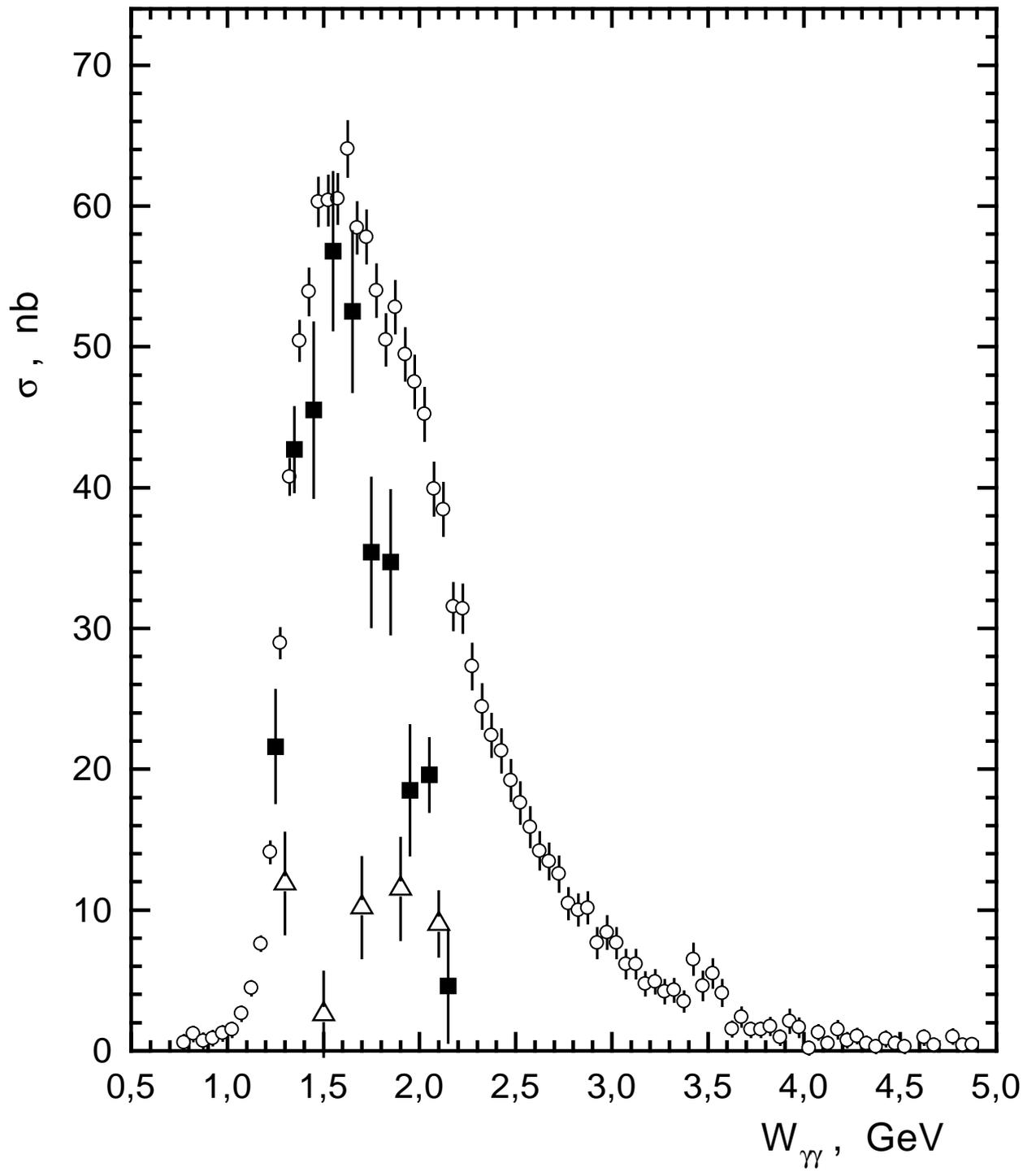
\end{center} {\bf Fig. 1} \,The L3 preliminary data on the $\gamma\gamma
\to\pi^+\pi^-\pi^+\pi^-$ cross section [8] (open circles) and the ARGUS data 
on the $(J^P,\,|J_z|)=(2^+,\,2)$ partial cross section for $\gamma\gamma\to
\rho^0\rho^0$ [10] (full squares) and $\gamma\gamma\to\rho^+\rho^-$ [11] (open
triangles). \end{document}